\documentclass[aps,twocolumn,showpacs,amsmath,superscriptaddress,amssymb,prc]{revtex4-2}
\usepackage{graphicx,here}
\usepackage{multirow}
\usepackage{tikz}
\usepackage{epstopdf}
\usepackage[percent]{overpic}
\usepackage{scrextend}
\usepackage{xcolor,xspace}
\usepackage[ulem=normalem]{changes}
\usepackage{hyperref}
\hypersetup{colorlinks=True,urlcolor=blue,linkcolor=blue,citecolor=blue,filecolor=black}

\colorlet{Changes@Color}{red}
\usepackage{dcolumn,color,footnote,bm,braket}
\usepackage{url,longtable,tabularx}
\usepackage{threeparttable}

\usepackage{fancybox}
\usetikzlibrary{matrix}

\begin{document}

\title{Collective-model description of shape coexistence and intruder states in cadmium isotopes based on a relativistic energy density functional}

\author{K.~Nomura}
\email{knomura@phy.hr}
\affiliation{Department of Physics, Faculty of Science, 
University of Zagreb, HR-10000 Zagreb, Croatia}

\author{K.~E.~Karakatsanis}
\affiliation{Department of Physics, Faculty of Science, 
University of Zagreb, HR-10000 Zagreb, Croatia}
\affiliation{Physics Department, Aristotle University of Thessaloniki, Thessaloniki GR-54124, Greece}

\date{\today}

\begin{abstract}
Low-energy structure of even-even $^{108-116}$Cd 
isotopes is analyzed using a collective model 
that is based on the nuclear density functional theory.  
Spectroscopic properties are computed by solving the 
triaxial quadrupole collective Hamiltonian, with 
parameters determined by the 
constrained self-consistent mean-field calculations 
within the relativistic Hartree-Bogoliubov method 
employing a universal energy density functional and 
a pairing force. 
The collective Hamiltonian reproduces 
the observed quadrupole phonon states 
of vibrational character, which are 
based on the moderately deformed 
equilibrium minimum 
in the mean-field potential energy surface. 
In addition, 
the calculation yields a low-lying excited 
$0^+$ band and a $\gamma$-vibrational band 
that are associated with 
a deformed local minimum close in energy to the 
ground state, consistently with the empirical 
interpretation of these bands as intruder bands. 
Observed energy spectra, 
$B(E2)$, and $\rho^2(E0)$ values are, in general, 
reproduced reasonably well. 
\end{abstract}

\maketitle

\section{Introduction}

Quadrupole collectivity  
is a basic, yet prominent, feature of 
nuclear structure, characterized by 
the (anharmonic) vibrations 
of a spherical nuclear surface or the rotations 
of an ellipsoidal deformed nuclear shape 
\cite{BM,RS,CasBook,IBM}. 
A collective vibrational spectrum is, in particular, 
observed in nearly spherical nuclei, 
and is interpreted  
in terms of the excitations of quadrupole phonons. 
The energy spectrum then consists 
of zero- ($0^+$), and one-phonon ($2^+$) states, 
followed by a two-phonon triplet
($4^+$, $2^+$, $0^+$) at twice the excitation 
energy of the one-phonon state, and so on. 
Classic examples of the 
vibrational energy spectrum have been known 
in stable nuclei near the 
proton $Z=50$ magic number, 
such as the even-even cadmium (Cd) 
\cite{goldhaber1955}, where 
the observed low-lying states 
indeed show features that resemble the 
quadrupole vibrational 
spectra predicted by the collective model 
of Bohr and Mottelson \cite{BM}.

Later experiments have revealed, however,  
in addition to the multiphonon states, 
extra $0^+$ and $2^+$ levels 
that are close in energy to the two-phonon 
triplet in even-even Cd \cite{cohen1961}. 
The appearance of these additional states 
is not explained in a vibrational 
picture, but rather implies that the 
pure quadrupole phonon interpretation 
of even-even Cd is untenable. 
In a spherical shell model the additional 
states were attributed to two-particle-two-hole 
(2p-2h) excitations of protons from 
above the $Z=50$ shell gap. 
Correlations between the valence 
protons and neutrons can be then 
so enhanced that the lowering 
of the intruder low-spin levels occurs 
\cite{federman77,heyde85,heyde1992,wood1992,heyde95,heyde2011}. 
The interpretation of the extra 
$0^+$ and $2^+$ states in Cd as  
2p-4h states was confirmed by the ($^3$He, $n$) 
reaction experiment \cite{fielding1977}. 
Furthermore, in the mean-field approximation 
\cite{bengtsson1987,bengtsson89,wood1992,andre00,heyde2011}
the normal and intruder states correspond to 
different minima on the potential 
energy surface defined in terms of the 
quadrupole deformations.

Along the chain of 
even-even Cd isotopes, the intruder bands 
have been shown to become lower in energy 
toward the middle of the major shell $N=66$ 
with the increasing number of valence neutrons. 
The structure of the even-even Cd has been 
studied by numerous experiments, most extensively,  
on stable isotopes with the mass $A$ = 106 
to $A$ = 116.  
An extensive list of the references to the 
related experimental studies 
is found in Ref.~\cite{nomura2018cd}. 
Recent reviews on the experimental and 
theoretical studies on the structure of the 
light and heavy Cd isotopes, as well as the neighboring 
isotopes in the tin (Sn) region, are given in 
Refs.~\cite{garrett2010,heyde2011,garrett2016,garrett2022}

Besides that, theoretical investigations of 
the even-even Cd have been performed from various 
perspectives. Large-scale shell model 
calculations have been carried out from the light 
($A$ $\approx$ 98) \cite{gorska1997,blazhev2004,ekstrom2009} 
up to the mass $A$ = 108 \cite{schmidt2017} Cd. 
As a more plausible approach that represents 
a drastic truncation 
of the shell model configuration space, 
calculations within an extended version of the 
interacting boson model (IBM) \cite{IBM} that 
takes into account the 2p-2h intruder excitations 
and configuration mixing 
between the normal (0p-0h) 
and intruder states have been 
carried out extensively 
\cite{heyde1982,heyde1992,deleze1993a,deleze1993b,heyde95,DeCoster96,Lehmann97,garrett2007,garrett2008,nomura2018cd,leviatan2018}. 
Alternative approaches are self-consistent mean-field 
(SCMF) methods \cite{RS} based on the nuclear density 
functional theory (DFT). Calculations within the 
symmetry-projected SCMF method using the Gogny-type 
\cite{Gogny,D1S} energy density functional (EDF) 
were performed to analyze the systematic behavior of 
the $2^+_1$ state of the even-even Cd 
in the entire $N$=50-82 major shell 
\cite{trodriguez2008}, to provide detailed 
descriptions of the spectroscopy of the 
$^{110,112}$Cd nuclei 
\cite{garrett2019,garrett2020}, 
and to describe in a systematic manner 
the low-energy structure of the 
even-even $^{98-130}$Cd nuclei in comparison 
with the updated experimental data \cite{siciliano2021}. 
A quadrupole collective Bohr Hamiltonian, 
derived from a microscopic framework of the 
adiabatic time-dependent Hartree-Fock-Bogoliubov 
method using a Skyrme force \cite{Skyrme}, 
was considered for $^{110-116}$Cd \cite{prochniak2012}.

Here we present an alternative theoretical 
description of the 
even-even $^{108-116}$Cd nuclei using 
the triaxial quadrupole collective 
Hamiltonian (QCH) that is based on the nuclear 
DFT. Within this theoretical scheme, 
parameters of the QCH 
are determined by using as microscopic 
inputs the solutions of the SCMF calculations 
based on a universal EDF and a pairing interaction. 
We shall identify, in most of the studied nuclei, 
low-energy collective bands that are 
associated with intruder bands as empirically 
suggested, 
and discuss their microscopic structures 
in connection with shape coexistence. 
In Sec.~\ref{sec:theory}, we give a brief description 
of the SCMF and QCH approaches. 
Results of the SCMF calculations 
are shown in Sec.~\ref{sec:mf}. 
In Sec.~\ref{sec:results}, we present the QCH 
results of the spectroscopic calculations, 
including the excitation energies, electric 
quadrupole and monopole transition rates, 
and detailed spectroscopy of $^{110,112}$Cd. 
Finally, Sec.~\ref{sec:summary} gives 
a summary of the main results.

\section{Theoretical framework\label{sec:theory}}

The first step in the theoretical procedure 
is to perform, for each nucleus, 
a set of the constrained SCMF calculations 
within the framework of the relativistic 
Hartree-Bogoliubov (RHB) method 
\cite{vretenar2005,niksic2011,DIRHB,DIRHBspeedup} 
employing the 
density-dependent point-coupling (DD-PC1) 
interaction \cite{DDPC1} and the separable 
pairing force of finite range developed 
in \cite{tian2009}. 
The constraints imposed in the SCMF calculations are 
on the expectation values of the 
mass quadrupole operators
\begin{align}
 \hat Q_{20}=2z^2-x^2-y^2 
\quad\text{and}\quad
 \hat Q_{22}=x^2-y^2 \; ,
\end{align}
which are related to the 
axially symmetric deformation $\beta$ 
and triaxiality $\gamma$ \cite{BM}, i.e., 
\begin{align}
\label{eq:bg}
& \beta=\sqrt{\frac{5}{16\pi}}\frac{4\pi}{3}\frac{1}{A(r_{0}A^{1/3})^{2}}
\sqrt{\braket{\hat{Q}_{20}}^{2}+2\braket{\hat{Q}_{22}}^{2}} \; , \\
& \gamma=\arctan{\sqrt{2}\frac{\braket{\hat{Q}_{22}}}{\braket{\hat{Q}_{20}}}} \; ,
\end{align}
with $r_0=1.2$ fm. 
The SCMF calculations are carried out 
in a harmonic oscillator basis, with the number of 
oscillator shells equal to 20. 
The strengths of the proton $V_p$ and neutron $V_n$
pairings are set equal, 
$V_0$ $\equiv$ $V_p$ $=$ $V_n$ $=$ 728 MeV fm$^3$, 
which have been obtained in Ref.~\cite{tian2009} 
so that the pairing gaps provided by the 
Gogny-D1S \cite{D1S} SCMF calculation 
are reproduced.

Quadrupole collective states are 
obtained as the solutions to the QCH. 
The parameters 
of the Hamiltonian are 
specified by using 
the results of the RHB calculations: 
the potential energy surfaces as 
functions of the $\beta$ and $\gamma$ 
deformations, and the single-particle solutions. 
The detailed accounts of this procedure 
are found in 
Refs.~\cite{niksic2009,niksic2011}. 
The collective Hamiltonian 
$\hat{H}_{\textnormal{coll}}$ 
is given as
\begin{align}
\label{eq:hamiltonian-quant}
\hat{H}_{\textnormal{coll}} 
= \hat{T}_{\textnormal{vib}}+\hat{T}_{\textnormal{rot}}
+V_{\textnormal{coll}} \; ,
\end{align}
with the vibrational kinetic energy:
\begin{align}
\label{eq:vib}
\hat{T}_{\textnormal{vib}} =
&-\frac{\hbar^2}{2\sqrt{wr}}
   \Biggl[\frac{1}{\beta^4}
   \Biggl(\frac{\partial}{\partial\beta}\sqrt{\frac{r}{w}}\beta^4
   B_{\gamma\gamma} \frac{\partial}{\partial\beta}
   \nonumber \\
&-\frac{\partial}{\partial\beta}\sqrt{\frac{r}{w}}\beta^3
   B_{\beta\gamma}\frac{\partial}{\partial\gamma}
   \Biggr)
   +\frac{1}{\beta\sin{3\gamma}}\Biggl(
   -\frac{\partial}{\partial\gamma} \sqrt{\frac{r}{w}}\sin{3\gamma}
\nonumber \\
&\times
B_{\beta \gamma}\frac{\partial}{\partial\beta}
    +\frac{1}{\beta}\frac{\partial}{\partial\gamma} \sqrt{\frac{r}{w}}\sin{3\gamma}
      B_{\beta \beta}\frac{\partial}{\partial\gamma}
   \Biggr)\Biggr] \; ,
\end{align}
rotational kinetic energy:
\begin{align}
\label{eq:rot}
\hat{T}_{\textnormal{rot}} =
\frac{1}{2}\sum_{k=1}^3{\frac{\hat{J}^2_k}{\mathcal{I}_k}} \; ,
\end{align}
and collective potential $V_{\textnormal{coll}}$. 
Note the operator $\hat{J}_k$ in Eq.~(\ref{eq:rot}) 
denotes the components of the angular momentum in
the body-fixed frame of a nucleus. 
The mass parameters 
$B_{\beta\beta}$, $B_{\beta\gamma}$, 
and $B_{\gamma\gamma}$ in (\ref{eq:vib}), 
and the moments of inertia 
$\mathcal{I}_k$ in (\ref{eq:rot}), 
are functions of the $\beta$ and $\gamma$ 
deformations, 
and are related to each other by 
$\mathcal{I}_k = 4B_k\beta^2\sin^2(\gamma-2k\pi/3)$. 
Two additional quantities in Eq.~(\ref{eq:vib}), i.e., 
$r=B_1B_2B_3$, and $w=B_{\beta\beta}B_{\gamma\gamma}-B_{\beta\gamma}^2 $,
determine the volume element in the collective space. 
The moments of inertia are
computed using the Inglis-Belyaev formula 
\cite{inglis1956,belyaev1961}, and the mass
parameters are calculated in the cranking 
approximation. 
The collective potential $V_{\textnormal{coll}}$ 
(\ref{eq:hamiltonian-quant}) is obtained by 
subtracting the zero-point energy corrections 
from the total RHB deformation energy.

The corresponding eigenvalue problem is solved using 
an expansion of eigenfunctions in terms
of a complete set of basis functions that depend on the 
deformation variables $\beta$ and
$\gamma$, and the Euler angles $\Omega=(\phi,\theta,\psi)$. 
The diagonalization of the Hamiltonian yields the excitation 
energies and collective
wave functions for each value of the total angular 
momentum and parity, 
that are used to calculate various physical observables. 
A virtue of using the QCH 
based on SCMF single-(quasi)particle 
solutions is the fact that the observables, 
such as electric quadrupole ($E2$) 
and monopole ($E0$) transition probabilities 
and spectroscopic quadrupole 
moments, are calculated in the full configuration 
space and there is no need for effective charges. 
Using the bare value of the proton charge in the
electric transition operators, 
the transition probabilities 
between eigenvectors of the QCH  
can be directly compared with 
spectroscopic data.

\begin{figure}
\begin{center}
\includegraphics[width=\linewidth]{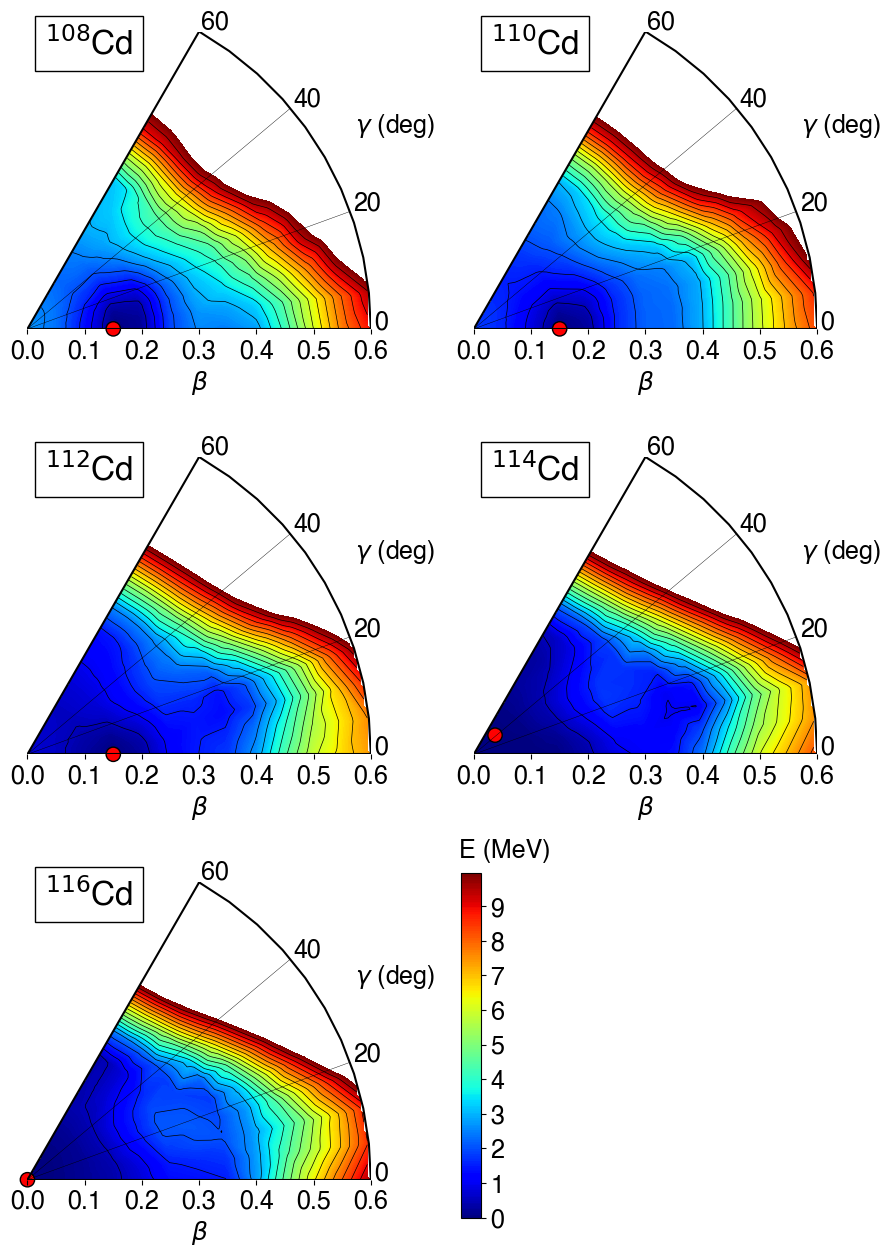}
\caption{Potential energy surfaces 
for the even-even $^{108-116}$Cd nuclei 
as functions of the triaxial quadrupole 
$\beta$-$\gamma$ deformations, 
computed by the constrained 
SCMF calculations within the RHB framework 
employing the 
interaction DD-PC1 and the separable pairing force 
of finite range. The total SCMF energies are plotted 
up to 10 MeV from the 
global minimum (indicated by the solid circle), 
and the energy difference between 
neighboring contours is 0.5 MeV. 
}
\label{fig:pes}
\end{center}
\end{figure}

\begin{figure*}
\begin{center}
\includegraphics[width=.7\linewidth]{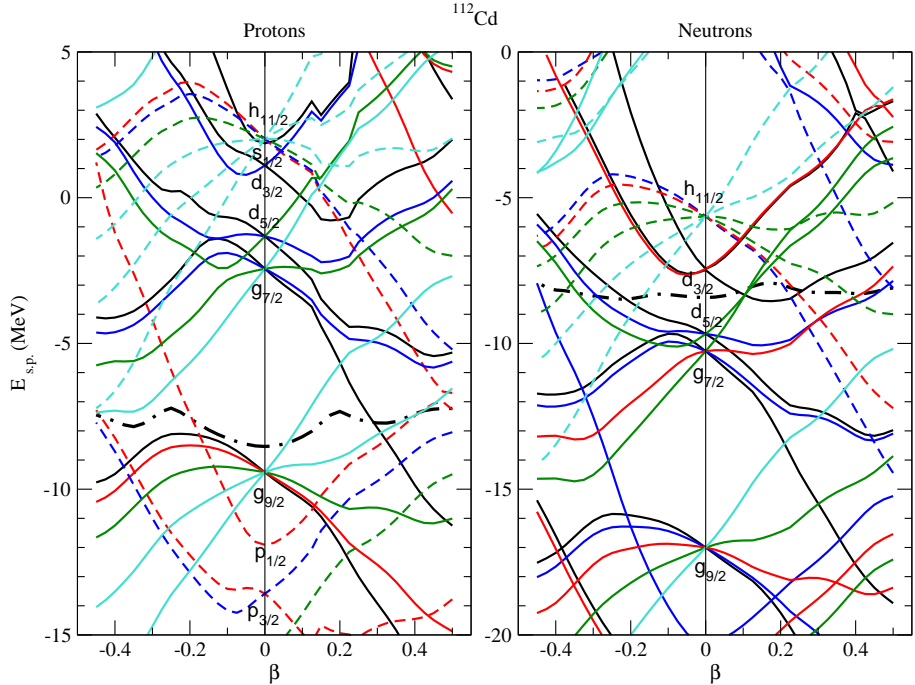}
\caption{Calculated single-particle energies 
for protons (left) and neutrons (right) 
for $^{112}$Cd as functions of the axial quadrupole 
deformation $\beta$. Dash-dotted curves 
represent Fermi energies.}
\label{fig:spe}
\end{center}
\end{figure*}

\begin{figure}
\begin{center}
\includegraphics[width=\linewidth]{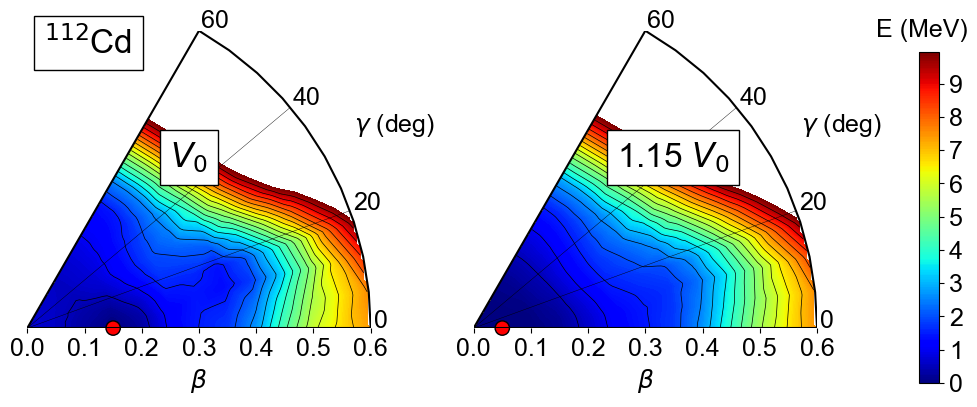}
\caption{Potential energy surfaces 
for $^{112}$Cd obtained from the constrained 
RHB method, with the 
pairing strength unchanged ``$V_0$'' (left) 
and increased by 15 \% for both 
protons and neutrons ``1.15$V_0$'' (right). 
}
\label{fig:pes-pair}
\end{center}
\end{figure}

\section{Mean-field results\label{sec:mf}}

Figure~\ref{fig:pes} shows 
the $\beta$-$\gamma$ triaxial potential 
energy surfaces for $^{108-116}$Cd 
calculated by using the constrained RHB method. 
The global minimum occurs 
at a weakly deformed prolate configuration 
$\beta$ $\approx$ 0.15 for $^{108,110,112}$Cd, 
and at a nearly spherical configuration 
for $^{114,116}$Cd. 
Besides the weakly deformed global minimum, 
in most of the nuclei 
two local minima with larger deformation 
$\beta$ $\gtrsim$ 0.3 
are obtained on both prolate 
($\gamma$ $=$ 0$^\circ$) and 
oblate ($\gamma$ $=$ 60$^\circ$) sides. 
In particular, a pronounced triaxial local 
minimum near the prolate axis, 
($\beta$, $\gamma$) $\approx$ (0.35, 12$^\circ$), 
which is close in energy to the global minimum, 
is suggested to occur for $^{112,114}$Cd.

The appearance of the minima in the potential 
energy surface is inferred from 
the behaviors of the single-particle levels  
near the Fermi energies. 
Figure~\ref{fig:spe} shows the 
single-particle energies for protons and neutrons 
for $^{112}$Cd as functions of the axial quadrupole 
deformation $\beta$, obtained as the 
SCMF solutions. 
In the proton single-particle spectra, 
near the Fermi energy 
(indicated by a dash-dotted curve 
in Fig.~\ref{fig:spe}), 
there is a gap within the range 
0.05 $\lesssim$ $\beta$ $\lesssim$ 0.3. 
The gap is produced essentially by the 
$g_{9/2}$ and $g_{7/2}$ orbitals, 
coming respectively from below and above 
the $Z$ = 50 major shell. 
In this range of deformation the global 
prolate minimum is obtained in the 
potential energy surface (see Fig.~\ref{fig:pes}). 
Another gap is seen in the single-proton spectra 
at $\beta$ $\approx$ 0.35, which is 
produced by the $g_{9/2}$ and $p_{1/2}$ from 
below the $Z$ = 50 major shell, and $g_{7/2}$ 
and $d_{5/2}$ from above. This corresponds 
to the local minimum that appears near the 
prolate axis 
($\beta$, $\gamma$) $\approx$ (0.35, 12$^\circ$). 
On the oblate side ($\beta<0$), 
yet another gap is visible 
in the interval 
$-0.4$ $\lesssim$ $\beta$ $\lesssim$ $-0.2$, 
created as a result 
of the lowering of the $g_{7/2}$ levels 
and the rising of the $g_{9/2}$, 
$p_{1/2}$, and $p_{3/2}$ ones. 
The gap is related to the oblate 
local minimum. 
The fact that the several 
energy gaps are obtained in the proton 
single-particle diagram 
conforms to the empirical interpretation 
that the observed extra low-spin states 
in Cd isotopes are attributed to 
particle-hole excitations of protons 
across the $Z$= 50 closed shell. 
Similarly, one could see 
in the neutron single-particle spectra 
(shown on the right hand side 
of Fig.~\ref{fig:spe}) 
energy gaps near the Fermi energy 
in those same ranges of the $\beta$ 
deformation at which the local minima occur 
in the potential energy surface. 
The gaps are, however, much less 
pronounced, i.e., the level density around 
the Fermi surface is 
much higher, than in the case of the 
single-proton spectra. 
It should be noted that the above 
argument, in terms of the appearance 
of the minima in the 
potential energy surface and the gaps 
in the single-particle levels, 
is made at the mean-field level, 
and provides only an approximate 
picture of low-lying states.

We further study the sensitivity of the calculations 
to the strengths of the proton $V_p$ and 
neutron $V_n$ pairing interactions. 
As an example, we show in Fig.~\ref{fig:pes-pair} 
the potential energy surfaces for $^{112}$Cd 
computed with the pairing strengths unchanged, 
i.e., $V_p$ $=$ $V_n$ $=$ 728 MeV fm$^3$ (= $V_0$), 
and increased by 15 \% for both protons and 
neutrons, i.e., 
$V_p$ $=$ $V_n$ $=$ $1.15V_0$ = 837 MeV fm$^3$. 
A comparison between the two  
surfaces in Fig.~\ref{fig:pes-pair} 
shows that, with the increased  
pairing strength, the global minimum shifts 
to the spherical side, $\beta$ $\approx$ 0.05, 
while the triaxial local minimum becomes 
much less pronounced. 
The same conclusion was reached in \cite{nomura2018cd} 
where the constrained Hartree-Fock 
plus BCS calculations for $^{112}$Cd 
using the Skyrme SLy6 force \cite{SLy} 
were employed as the input to build the IBM 
Hamiltonian with configuration mixing. 
In the following, we mainly discuss results 
with the original pairing strength in the 
RHB calculations ($V_p$ $=$ $V_n$ $=$ $V_0$), 
while the dependence of the spectroscopic 
properties on the pairing strengths 
will also be analyzed.

\begin{figure*}[ht]
\begin{center}
\includegraphics[width=\linewidth]{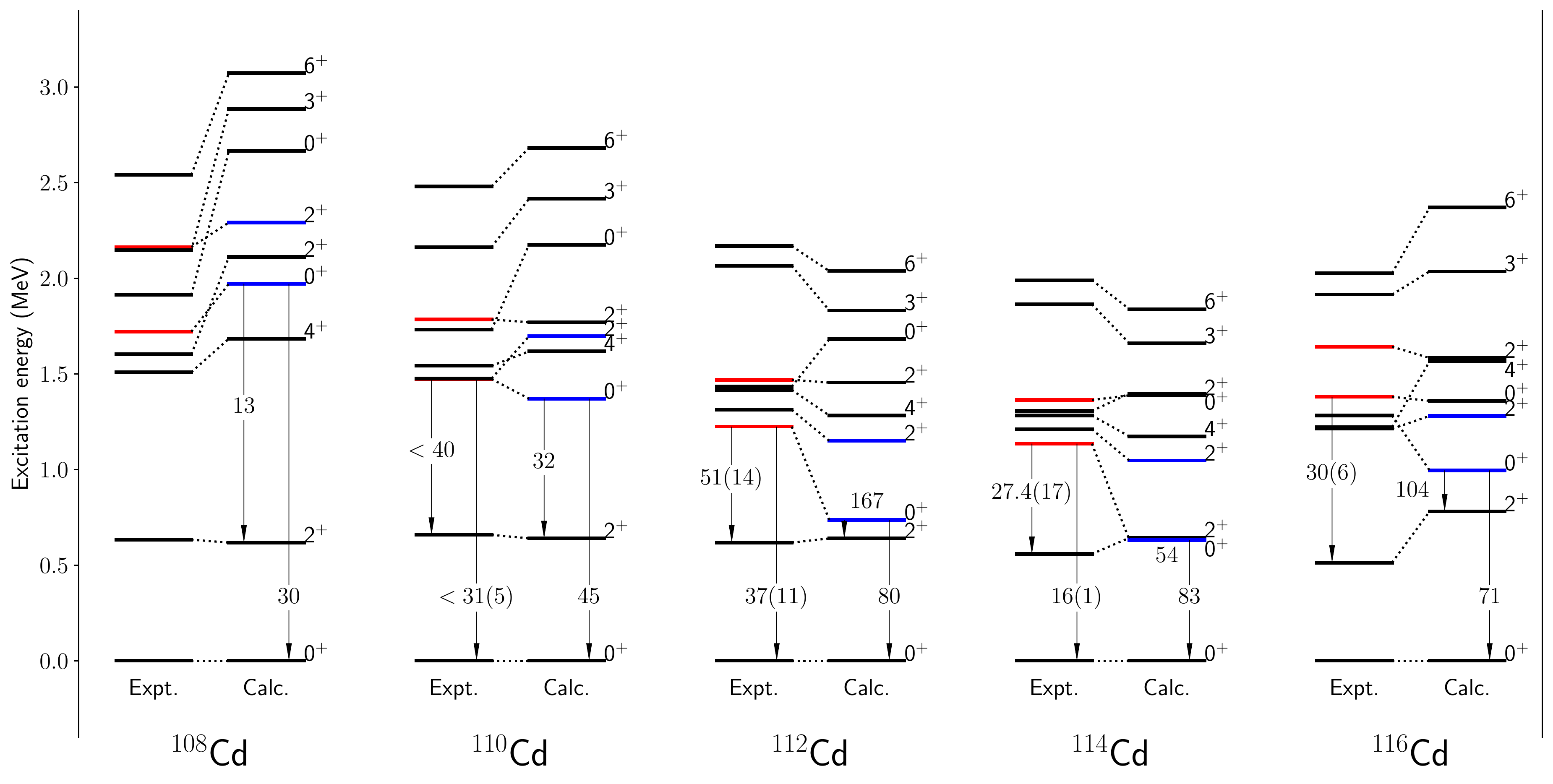}
\caption{
Comparison between the calculated and experimental 
low-energy spectra for the even-even $^{108-116}$Cd 
isotopes. 
Numbers with arrows from the $0^+_2$ level 
to $2^+_1$ and $0^+_1$ represent 
the $B(E2; 0^+_2 \to 2^+_1)$ values in Weisskopf units (W.u.), 
and the $\rho^2(E0; 0^+_2 \to 0^+_1)\times 10^3$ 
values, respectively. 
Only for $^{114}$Cd, the calculated $0^+_2$ level 
is below the $2^+_1$ one, and therefore 
the $B(E2; 2^+_1 \to 0^+_2)$ value is given. 
The experimental data are taken from 
Refs.~\cite{kibedi2005,garrett2016,garrett2019,garrett2020,data}. 
The experimental levels 
that are highlighted in color red represent 
the suggested intruder states, while 
the corresponding theoretical levels 
are in color blue.
}
\label{fig:level}
\end{center}
\end{figure*}

\begin{figure}
\begin{center}
\includegraphics[width=\linewidth]{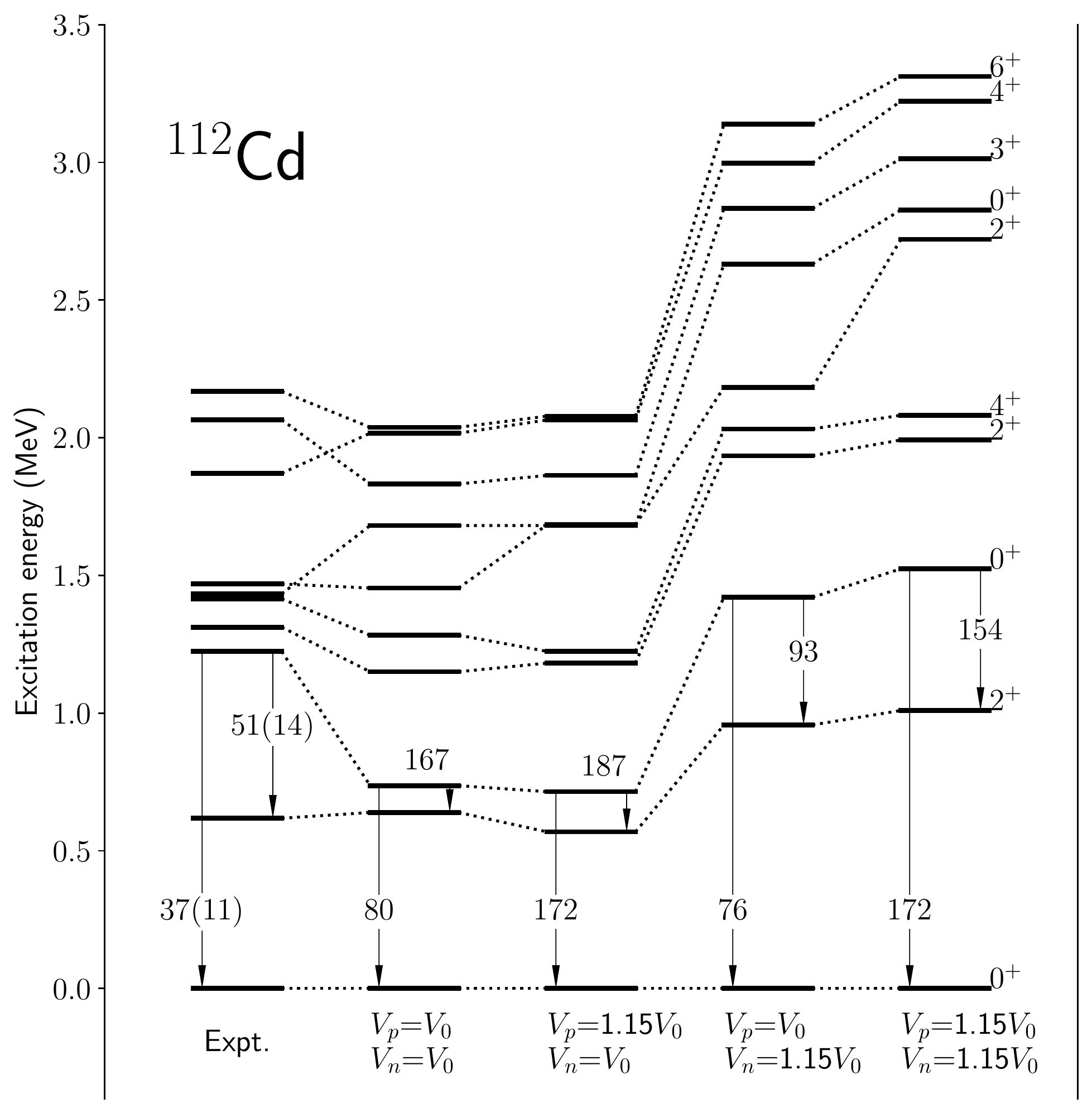}
\caption{Predicted excitation spectra 
for $^{112}$Cd with the 
pairing strength 
unchanged ($V_p$ = $V_0$, $V_n$ = $V_0$), 
increased by 15 \% 
for protons only
($V_p$ = $1.15V_0$, $V_n$ = $V_0$), 
for neutrons only
($V_p$ = $V_0$, $V_n$ = $1.15V_0$), 
and for both protons and neutrons 
($V_p$ = $1.15V_0$, $V_n$ = $1.15V_0$) 
in the RHB calculations. 
$B(E2;0^+_2 \to 2^+_1)$ (in W.u.) and 
$\rho^2(E0; 0^+_2 \to 0^+_1)\times 10^3$ values 
are also shown. 
The experimental data are 
taken from \cite{garrett2019}.
}
\label{fig:pair-level}
\end{center}
\end{figure}

\begin{figure*}
\begin{center}
\includegraphics[width=\linewidth]{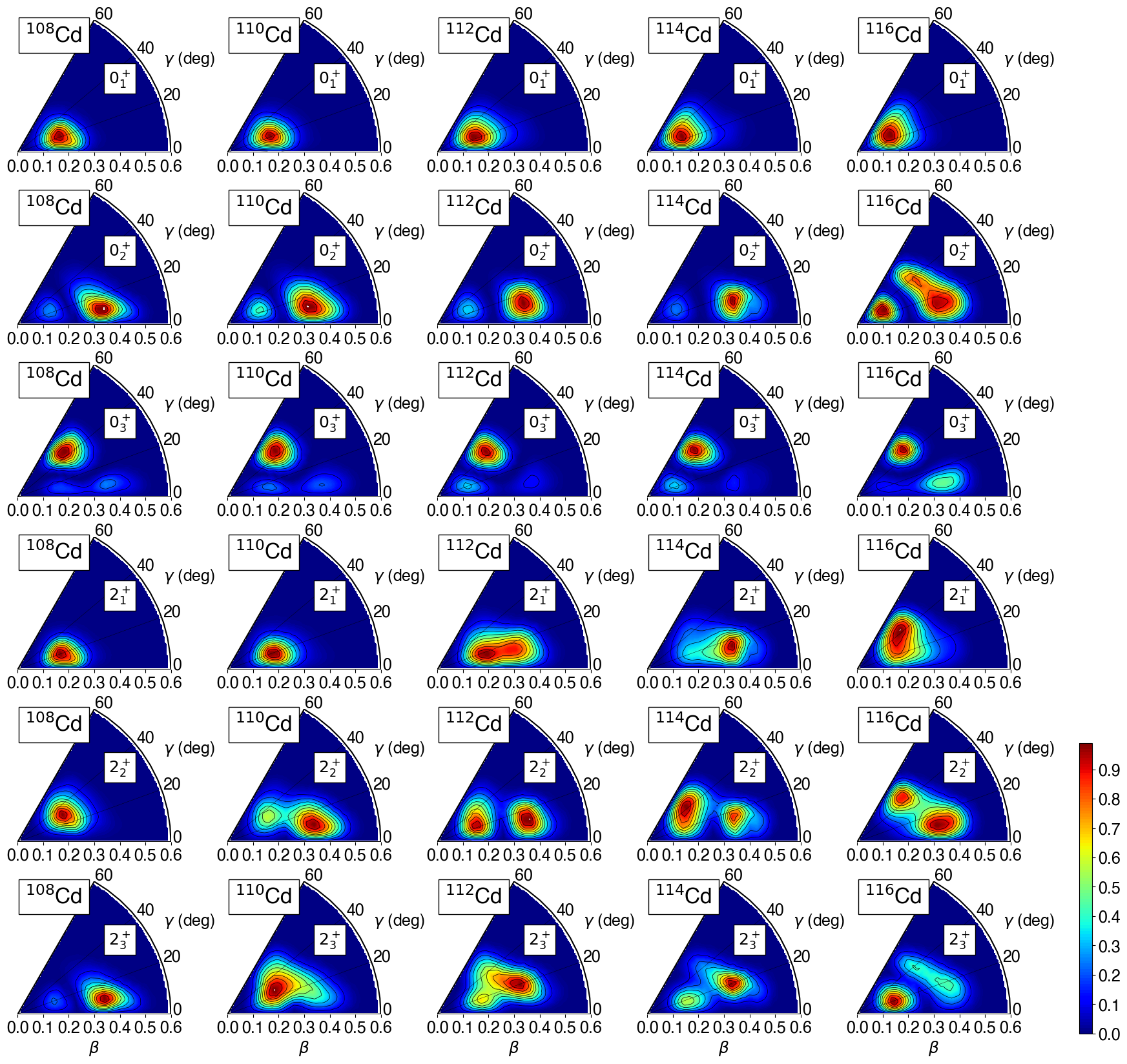}
\caption{
Distributions of collective wave functions 
for the $0^+_{1,2,3}$, and $2^+_{1,2,3}$ states 
of $^{108-116}$Cd 
within the ($\beta$, $\gamma$) plane.}
\label{fig:cwf}
\end{center}
\end{figure*}

\section{Spectroscopic results\label{sec:results}}

Figure~\ref{fig:level} shows the 
calculated low-energy excitation spectra, 
$B(E2; 0^+_2 \to 2^+_1)$ and 
$\rho^2(E0; 0^+_2 \to 0^+_1)$ transition probabilities 
for the even-even nuclei $^{108-116}$Cd. 
Experimental data are taken from 
Refs.~\cite{kibedi2005,garrett2016,garrett2019,garrett2020,data}. 
For $^{108-114}$Cd, the $0^+_2$ 
state has been empirically suggested to 
be the bandhead of the intruder band associated with 
the proton 2p-2h excitations \cite{gade2002a}. 
As for $^{116}$Cd, the $0^+_3$ state has been 
identified as the lowest intruder state. 
For all five nuclei, the $2^+_3$ state is 
attributed to the first excited state of the 
intruder band by experiments. 
In the present RHB+QCH calculation, 
as shown later, the 
$0^+_2$ and $2^+_2$ states are considered 
as the corresponding intruder states. 
Only for $^{108}$Cd, the $2^+_3$ state is 
here suggested to be of intruder nature.

The present calculation reproduces 
the energies of the normal, i.e., phonon-like states 
($2^+_1$, $4^+_1$, $6^+_1$, $3^+_1$, $2^+_2$, 
and $0^+_3$) fairly well. 
The observed intruder $0^+$ and $2^+$ 
states gradually decrease in energy and become lowest 
at $^{114}$Cd corresponding to the midshell $N=66$. 
The measured $E2$ and $E0$ transitions from 
the intruder $0^+_\text{intr}$ state, i.e., 
$B(E2; 0^+_\text{intr} \to 2^+_1)$
and 
$\rho^2(E0; 0^+_\text{intr} \to 0^+_1)$, 
also increase toward $^{114}$Cd. 
The RHB+QCH calculation gives similar 
systematic behaviors of these quantities, 
but underestimates the intruder $0^+$ and $2^+$ 
level energies for $^{112,114}$Cd significantly. 
Moreover, the predicted 
$B(E2; 0^+_2 \to 2^+_1)$
and $\rho^2(E0; 0^+_2 \to 0^+_1)$ 
values are generally a factor of two to three 
larger than the experimental values.

Figure~\ref{fig:pair-level} compares 
the excitation energies for $^{112}$Cd, 
obtained with the pairing strengths 
unchanged 
($V_p$, $V_n$) = ($V_0$, $V_0$), 
increased by 15 \% for protons only 
($V_p$, $V_n$) = (1.15$V_0$, $V_0$), 
for neutrons only 
($V_p$, $V_n$) = ($V_0$, 1.15$V_0$), 
and for both protons and neutrons 
($V_p$, $V_n$) = (1.15$V_0$, 1.15$V_0$), 
in the RHB calculations. 
It is seen that the increase in the proton pairing 
does not have any notable effect on energy spectra, 
but enhances the $B(E2; 0^+_2 \to 2^+_1)$
and $\rho^2(E0; 0^+_2 \to 0^+_1)$ transition probabilities. 
On the other hand, if the neutron pairing strength is 
increased, the $0^+_2$ level is raised to be 
closer in energy to the experimental counterpart. 
In this case, however, the whole energy spectrum 
becomes stretched, and overestimates the experimental 
spectrum. 
The finding that the change in the 
proton pairing strength does not have 
notable influence on the spectra reflects  
the fact that the studied Cd nucleus 
is close to the 
proton $Z=50$ major shell closure, around which 
the number of valence protons 
($Z_\text{val}$ = 2) is not 
large enough to make a sizable 
contribution to the low-energy spectra. 
On the other hand, the increase in the 
neutron pairing appears to have a more significant 
effect on the low-lying levels than 
that for the proton pairing. 
This is probably because, as the nucleus is 
close to the middle of the 
neutron major shell $N$ = 50-82, 
there are more valence neutrons 
($N_\text{val}$ = 14 for $^{112}$Cd), 
which are supposed to play a more dominant 
role in low-lying states.

Furthermore, the fact that increasing the 
pairing strength generally raises the energy 
levels, as one observes in Fig.~\ref{fig:pair-level}, 
is also anticipated from the comparison 
of the potential energy surfaces 
(see Fig.~\ref{fig:pes-pair}), 
which have been obtained with the 
pairing strength increased and unchanged in the 
constrained RHB calculations. 
With the increased pairing, the energy 
surface indicates a less deformed shape; hence 
the energy spectrum should become more of 
vibrational character.

To provide an insight into the intruder 
nature of the predicted excited states, 
in Fig.~\ref{fig:cwf} we show the distributions of 
the collective wave functions in the 
$\beta$-$\gamma$ plane for the $0^+_{1,2,3}$ 
and $2^+_{1,2,3}$ states of $^{108-116}$Cd. 
The wave function of the $0^+_1$ ground state 
in all the studied nuclei 
is sharply peaked at weakly deformed 
(triaxial) configurations 
($\beta$, $\gamma$) $\approx$ (0.15, 20-40$^\circ$), 
the coordinate corresponding to the 
weakly deformed global minimum in the 
potential energy surface (see Fig.~\ref{fig:pes}). 
The $0^+_2$ wave function shows 
a sharp peak at larger deformation, 
($\beta$, $\gamma$) $\approx$ 
(0.35, 10$^\circ$). 
This deformation configuration 
corresponds to the local 
minimum near the prolate axis in the 
potential energy surfaces, hence the 
$0^+_2$ is here assigned to be 
the bandhead of the intruder band. 
Two major peaks are obtained for 
the $0^+_2$ wave function for $^{116}$Cd: 
($\beta$, $\gamma$) $\approx$ 
(0.35, 15$^\circ$) and  (0.1, 30$^\circ$). 
The one at ($\beta$, $\gamma$) $\approx$ 
(0.35, 15$^\circ$) is also spread along 
the $\gamma$ deformation. Hence, 
a considerable amount of shape mixing 
is expected to be present 
in the $0^+_2$ state of $^{116}$Cd. 
This is related to the fact that 
the energy surface for $^{116}$Cd 
is considerably soft in the $\gamma$ direction.

The $0^+_3$ wave function distributions, shown 
in the third row of Fig.~\ref{fig:cwf}, 
generally exhibit 
a major peak on the oblate side, 
corresponding to the local oblate minimum or 
saddle point. 
The $2^+_1$ collective wave function for 
$^{108-112}$Cd shows a similar distribution 
pattern to $0^+_1$, as it is peaked at 
($\beta$, $\gamma$) $\approx$ (0.15, 20$^\circ$). 
For the $2^+_1$ states of 
$^{114,116}$Cd, however, 
the peak appears at 
($\beta$, $\gamma$) $\approx$ 
(0.35, 15$^\circ$) for $^{114}$Cd and 
(0.2, 40$^\circ$) for $^{116}$Cd, 
at variance with the distribution patterns 
of the respective 
$0^+_1$ collective wave functions.  
The $2^+_1$ wave function for 
$^{112-116}$Cd is also spread 
over wider regions in the ($\beta$, $\gamma$) 
plane than for $^{108,110}$Cd. 
This implies that the mixing between 
different shape configurations 
is present already in the normal state 
$2^+_1$ of $^{112-116}$Cd. 
The calculated $2^+_2$ states for $^{110-116}$Cd, 
and $2^+_3$ state for $^{108}$Cd can be 
associated with 
the $2^+$ members of the observed intruder states, 
based on the fact that in the present 
calculation these nonyrast 
$2^+$ states are shown to exhibit 
a particularly strong $E2$ transition 
to the $0^+_2$ state (see Fig.~\ref{fig:level}). 
Indeed, for most of these nuclei, the collective 
wave function distribution gives two peaks, which  
have a large overlap with the $0^+_2$ wave function. 
Furthermore, the calculation suggests 
the $2^+_3$ ($2^+_2$) state for $^{110-116}$Cd 
($^{108}$Cd) to be the 
bandhead of the lowest $\gamma$-vibrational 
or $K=2^+$ band. 
This interpretation is based on the dominance 
of the $K=2$ components in these states. 
The corresponding collective wave functions 
are indeed peaked at the triaxial region with 
$\gamma$ $\approx$ 30$^\circ$.

\begin{figure*}[ht]
\begin{center}
\includegraphics[width=\linewidth]{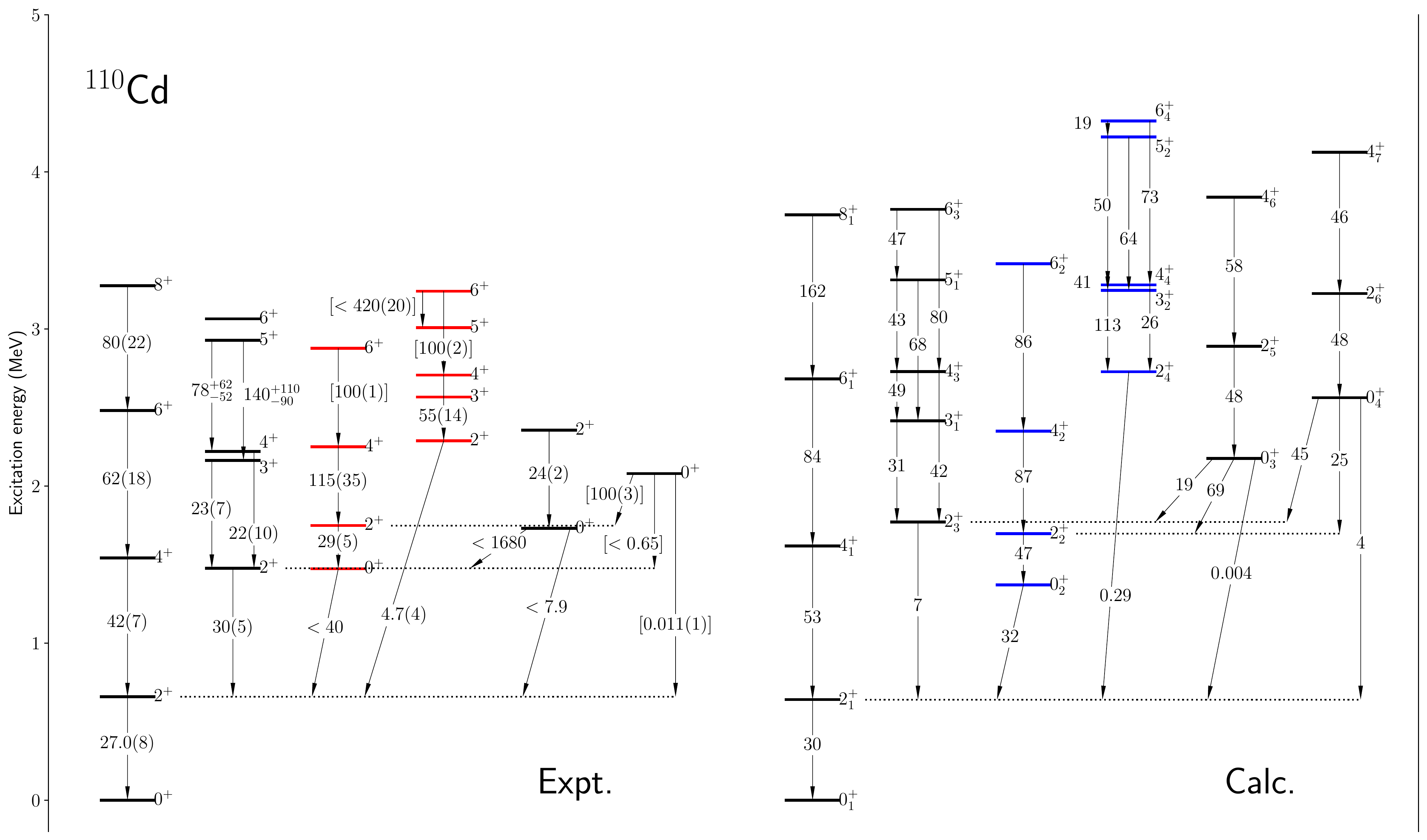}
\caption{Calculated and experimental \cite{garrett2019}
excitation spectra and $B(E2)$ transition rates (in W.u.) 
for the $^{110}$Cd nucleus. 
Following the notations 
in Ref.~\cite{garrett2019}, the experimental 
$B(E2)$ values in parentheses stand for 
relative transition strengths. 
The experimental levels 
that are highlighted in color red represent 
the suggested intruder states, and 
the corresponding theoretical levels 
are in color blue.
}
\label{fig:cd110}
\end{center}
\end{figure*}

\begin{figure*}[ht]
\begin{center}
\includegraphics[width=\linewidth]{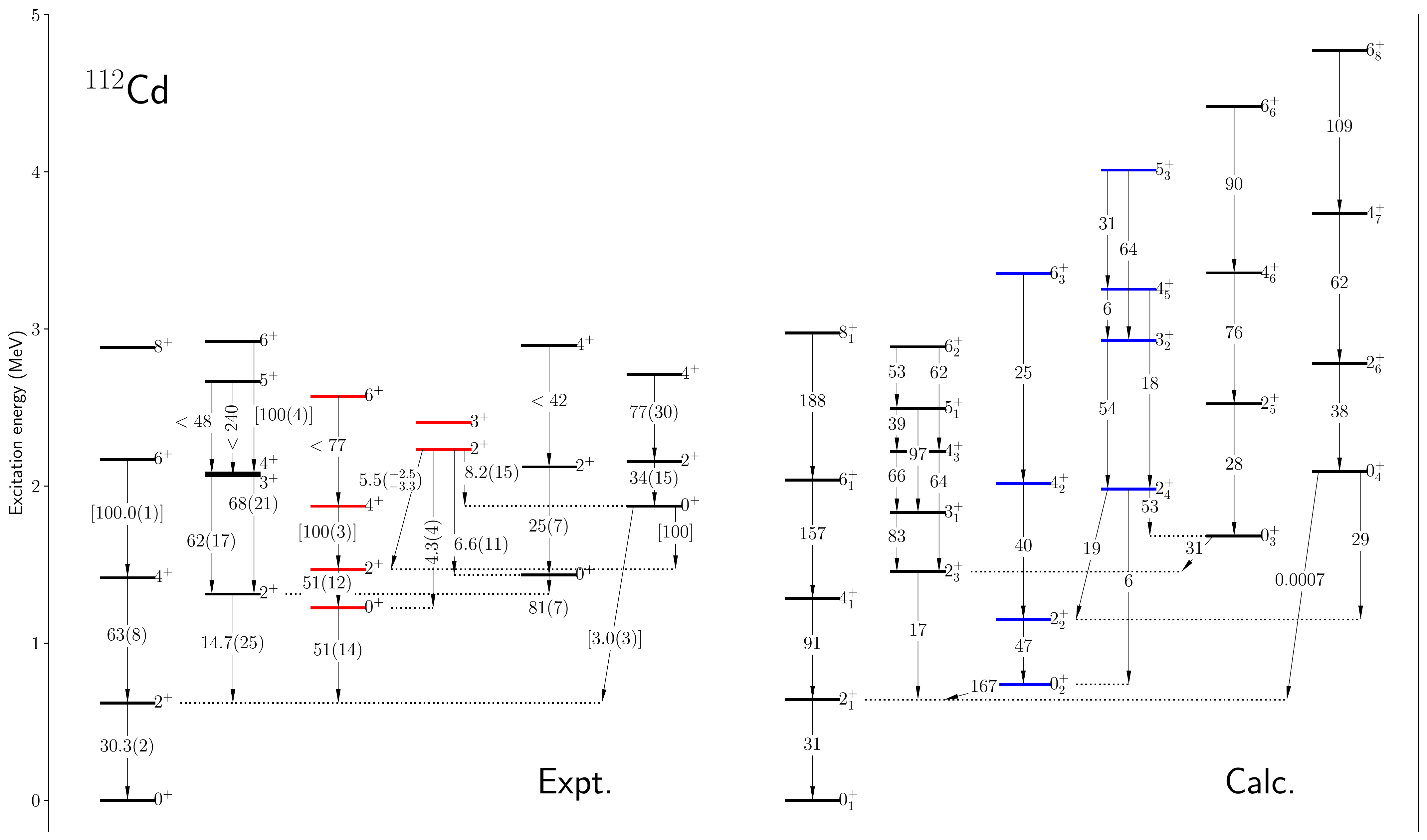}
\caption{Same as Fig.~\ref{fig:cd110}, but for the 
$^{112}$Cd nucleus.}
\label{fig:cd112}
\end{center}
\end{figure*}

Figure~\ref{fig:cd110} shows 
the calculated low-energy band structure 
including $B(E2)$ transition 
probabilities for $^{110}$Cd, 
in comparison with the 
experimental data \cite{garrett2019}. 
For the theoretical energy spectra, 
states are classified into 
the ground-state, lowest three $K=0^+$, 
and lowest two $K=2^+$ bands 
according to the dominant $E2$ transitions 
within the band and to the similarity 
in the fractions 
of the $K$ $=$ 0, 2, and 4 components. 
The observed low-energy spectra have 
multiphonon structure typical of 
vibrational nuclei, that is, 
the approximate 
one-phonon state $2^+_1$, 
two-phonon triplet 
($4^+_1$, $2^+_2$, $0^+_3$) at twice its energy, 
and three-phonon quintet
($6^+_1$, $4^+_2$, $3^+_1$, $2^+_5$, $0^+_4$) 
at three times the $2^+_1$ energy. 
The $\Delta I$ = 2 band built on the excited 
$0^+_2$ state has been assigned to be an 
intruder band by experiment \cite{garrett2019}. 
The second $K=2^+$ band, which is based upon 
the $2^+_4$ state, has also been found to be 
the intruder $\gamma$ band experimentally. 
In the present 
calculation, a phonon-like level structure 
appears as  the closely lying 
($4^+_1$, $2^+_3$, $0^+_3$), and 
($6^+_1$, $4^+_3$, $3^+_1$, $2^+_5$, $0^+_4$) 
states. 
The intruder bands that can be identified 
by the RHB+QCH calculation are those based 
on the $0^+_2$ and $2^+_4$ states. 
The calculation reproduces the $K=0^+_2$ 
intruder band rather well, except that the 
the energy level of the $6^+$ member 
is overestimated. 
A large transition strength 
$B(E2; 0^+_2 \to 2^+_1)$ $=$ 32 W.u. is 
here obtained, being consistent with the 
experimental value $<$40 W.u. 
The intruder $\gamma$ (or second $K=2^+$) 
band is, however, 
calculated to be rather high and stretched in 
energy by the RHB+QCH, as compared with the data. 
States in the calculated 
second $K=2^+$ band are connected by the strong 
$\Delta I$ = 1, as well as $\Delta I$ = 2, inband 
$E2$ transitions. 
The calculated $K=0^+_{3}$ ($0^+_4$) band is  
slightly higher than 
the observed one, and exhibits 
large interband $E2$ transition probabilities 
$B(E2; 0^+_{K=0^+_3} \to 2^+_{K=2^+})$ = 19 W.u.,
and $B(E2; 0^+_{K=0^+_3} \to 2^+_{K=0^+_2})$ = 69 W.u. 
[$B(E2; 0^+_{K=0^+_4} \to 2^+_{K=2^+})$ = 45 W.u. 
and $B(E2; 0^+_{K=0^+_4} \to 2^+_{K=0^+_2})$ = 25 W.u.]. 
Here $2^+_{K=2^+}$ and $2^+_{K=0^+_2}$ 
denote the $2^+_3$ bandhead 
of the lowest $K=2^+$ and the 
$2^+$ member of the $K=0^+_2$ band, respectively. 
The spectroscopic quadrupole moment 
for the $2^+_1$ state is calculated to be 
$Q(2^+_1)$ $=$ $-0.53$ $e$b, 
slightly larger in magnitude 
than the experimental 
value $-0.40(3)$ $e$b \cite{garrett2019}.

Figure~\ref{fig:cd112} shows the energy 
spectra for $^{112}$Cd. 
The whole energy spectrum, both theoretical 
and experimental, appears to be rather compressed 
in comparison with $^{110}$Cd. 
The present RHB+QCH calculation 
reproduces the observed energy levels 
reasonably well, apart from the 
fact that the $0^+_2$ band 
is obtained at much lower energy and 
is more stretched with increasing spin 
than in experiment. 
Similarly to the case of $^{110}$Cd, 
in addition to the phonon-like states grouped 
into approximate multiplets 
($4^+_1$, $2^+_2$, $0^+_3$), 
($6^+_1$, $4^+_3$, $3^+_1$, $2^+_4$, $0^+_4$), 
$\ldots$, etc., 
the states belonging to the band built on the 
$0^+_2$ state, together with the 
additional $2^+_6$ and $3^+_2$ states, 
have been experimentally suggested to be 
intruder states arising from the 
proton 2p-2h excitations. 
The corresponding two intruder bands in 
the present calculation are the ones 
starting from the $0^+_2$ and $2^+_4$ states. 
A possible reason 
for the $K=0^+_2$ band being calculated 
to be significantly low in energy 
is that the local triaxial minimum at 
($\beta$, $\gamma$) $\approx$ (0.35, 12$^\circ$) 
in the potential energy surface 
is too pronounced (see Fig.~\ref{fig:pes}). 
The states belonging to the $0^+_2$ band are 
mainly constructed from this local minimum. 
The fact that the calculated transition rate 
$B(E2; 0^+_2 \to 2^+_1)$ = 167 W.u., is 
a factor of three greater than 
the measured one ($51 \pm 14$ W.u.) further 
corroborates the occurrence of strong shape mixing. 
The predicted $0^+_3$ and $0^+_4$ excitation 
energies are, however, close to the experimental ones. 
Their $E2$ selection rules also follow 
what are observed experimentally: 
the large transition probability 
from the $0^+_3$ state to the 
$2^+_{K=2^+}$ bandhead 
[$B(E2; 0^+_3 \to 2^+_3)$ = 31 W.u.], 
and the dominance of 
the $0^+_{K=0^+_4} \to 2^+_{K=0^+_2}$ $E2$ 
transition over the $0^+_{K=0^+_4} \to 2^+_{g.s.}$ one. 
The $Q(2^+_1)$ moment of $^{112}$Cd 
is calculated to be 
$-0.68$ $e$b, which is larger 
in magnitude than 
the experimental value $Q(2^+_1)$ $=$ $-0.38$ $e$b 
\cite{garrett2019}, 
as in the $^{110}$Cd case discussed earlier. 

Recent theoretical calculations 
for $^{110,112}$Cd 
\cite{garrett2019,garrett2020} within the 
symmetry conserving configuration 
mixing (SCCM) method using 
the Gogny force generally overestimated the 
energy levels of the observed excited $0^+$ 
states, whereas in the present RHB+QCH 
calculation these $0^+$ energy levels, 
particularly the one for the 
second $0^+$ state, are predicted to 
be much lower. In addition, 
the same Gogny plus SCCM calculations provided the 
ground-state band for both the $^{110,112}$Cd nuclei 
that is rather stretched in energy 
with respect to the one obtained in the 
present calculation. 
It should be noted that, in solving 
the collective Hamiltonian in the 
present study, we do not make 
any adjustment of the cranking moment of inertia, 
e.g., increase of it by 30-40 \% to reproduce 
experiment, that is often considered 
in the literature.

The IBM calculation, using the boson 
Hamiltonian with partial dynamical symmetry (PDS)
breaking \cite{leviatan2018} and taking 
into account the 
configuration mixing between normal 
and intruder states, was also 
carried out to study the possible 
breakdown of the vibrational structure 
of $^{110}$Cd. 
By virtue of introducing the PDS, 
the IBM calculation 
obtained additional 
low-lying states close in energy to the 
normal vibrational states, 
which correspond to the empirically 
suggested intruder states. 
The calculated energy spectra and $B(E2)$ 
rates reported in that reference 
fit very well the experimental ones, 
while the parameters for the boson 
Hamiltonian and effective charges for 
the quadrupole operator were there 
determined by a phenomenological 
adjustment to the empirical data.

\section{Summary\label{sec:summary}}

In summary, we have analyzed the structure of the 
even-even $^{108-116}$Cd isotopes within the framework 
of a general collective model that is based on the 
nuclear density functional theory. 
Parameters of the 
triaxial quadrupole collective Hamiltonian, 
i.e., deformation-dependent mass parameters, 
moments of inertia, and collective potential, 
are determined by using as microscopic inputs 
the solutions to the constrained mean-field 
calculations within the relativistic 
Hartree-Bogoliubov approach. 
The mean-field results 
for the near midshell nuclei $^{112,114}$Cd 
indicate coexistence of normal states that are 
associated with a weakly deformed prolate 
or nearly spherical global minimum, and intruder 
states constructed from a more deformed, 
nearly prolate 
triaxial, local minimum.

Observed low-energy spectra, 
$B(E2)$, and $\rho^2(E0)$ values 
have been described reasonably well by the 
collective Hamiltonian. 
The present spectroscopic calculation 
produced a low-energy $0^+_2$ band and an additional 
$\gamma$-vibrational band which correspond to a 
triaxial local minimum in the potential energy 
surface, consistent with the empirical 
assignment of these bands as intruder bands. 
The calculation has reproduced an observed decrease 
of the intruder bands toward the midshell $N=66$ 
as a function of nucleon number. 
The intruder interpretation of the 
$0^+_2$ band associated 
with the local minimum has been 
further confirmed by distributions of the 
collective wave functions in 
the $\beta$-$\gamma$ plane. 
For $^{112,114}$Cd, however, 
the QCH underestimates the observed intruder 
bands significantly, and results in 
the too large $B(E2; 0^+_2 \to 2^+_1)$ and 
$\rho^2(E0; 0^+_2 \to 0^+_1)$ values, 
even though the approximate multiplets 
of energy levels typical of the phonon-like 
normal states were reasonably described. 
The discrepancy in the intruder bands 
reflects the too pronounced 
triaxial local minimum, which reflects, 
to a large extent, properties of the 
energy density functional employed in this study. 
We have also studied the sensitivity of the 
results to the pairing strengths in the case 
of $^{112}$Cd. 
It has been shown that with 
the increase in the strength of the 
(especially neutron) 
pairing, the local minima disappear 
in the potential 
energy surface, leading to a less deformed 
structure, and the corresponding energy 
levels are significantly raised.

\acknowledgments
The authors are grateful to Antonio Bjel\v ci\'c for helping 
them with implementation of the modified RHB solver. 
This work is financed within the Tenure Track Pilot Programme of 
the Croatian Science Foundation and the \'Ecole Polytechnique 
F\'ed\'erale de Lausanne, and the Project TTP-2018-07-3554 Exotic 
Nuclear Structure and Dynamics, with funds of the Croatian-Swiss 
Research Programme.

\bibliography{refs}

\end{document}